\documentclass[aps,prx,twocolumn,showpacs,amsmath,amssymb,superscriptaddress,floatfix]{revtex4-1}

\usepackage{graphicx}
\usepackage{dcolumn}
\usepackage{bm}
\usepackage{amssymb}
\usepackage{booktabs}
\usepackage{color}
\usepackage{microtype}
\usepackage[T1]{fontenc}
\usepackage{lmodern} 
\usepackage{enumitem}
\usepackage[colorlinks,plainpages=false,linkcolor=blue,urlcolor=blue,citecolor=blue,pdfpagemode=UseNone,pdfstartview=FitBH]{hyperref}

\usepackage{sidecap}

\renewcommand{\tablename}{Table}
\makeatletter\renewcommand{\fnum@table}[1]{\tablename~\thetable.}\makeatother

\newcommand{\RNO}{$R$NiO$_3$}
\newcommand{\NNO}{NdNiO$_3$}
\newcommand{\tmi}{$T_{\mathrm{MIT}}$}

\newcommand{\vb}[1]{\mathbf{#1}}

\begin{document}

\title{Site-selective Probe of Magnetic Excitations in Rare-earth Nickelates using \\ Resonant Inelastic X-ray Scattering}

\author{Y.~Lu}
\affiliation{Max-Planck-Institut f\"ur Festk\"orperforschung, Heisenbergstrasse~1, 70569 Stuttgart, Germany}

\author{D.~Betto}
\affiliation{European Synchrotron Radiation Facility, 71 Avenue des Martyrs, Grenoble F-38043, France}

\author{K.~F\"ursich}
\affiliation{Max-Planck-Institut f\"ur Festk\"orperforschung, Heisenbergstrasse~1, 70569 Stuttgart, Germany}

\author{H.~Suzuki}
\affiliation{Max-Planck-Institut f\"ur Festk\"orperforschung, Heisenbergstrasse~1, 70569 Stuttgart, Germany}

\author{H.-H.~Kim}
\affiliation{Max-Planck-Institut f\"ur Festk\"orperforschung, Heisenbergstrasse~1, 70569 Stuttgart, Germany}

\author{G.~Cristiani}
\affiliation{Max-Planck-Institut f\"ur Festk\"orperforschung, Heisenbergstrasse~1, 70569 Stuttgart, Germany}

\author{G.~Logvenov}
\affiliation{Max-Planck-Institut f\"ur Festk\"orperforschung, Heisenbergstrasse~1, 70569 Stuttgart, Germany}

\author{N.\,B.~Brookes}
\affiliation{European Synchrotron Radiation Facility, 71 Avenue des Martyrs, Grenoble F-38043, France}

\author{E. Benckiser}
\affiliation{Max-Planck-Institut f\"ur Festk\"orperforschung, Heisenbergstrasse~1, 70569 Stuttgart, Germany}

\author{M.\,W.~Haverkort}
\affiliation{Institut f\"ur Theoretische Physik, Universit\"at Heidelberg, Philosophenweg 19, 69120 Heidelberg, Germany}

\author{G.~Khaliullin}
\affiliation{Max-Planck-Institut f\"ur Festk\"orperforschung, Heisenbergstrasse~1, 70569 Stuttgart, Germany}

\author{M. Le Tacon}
\affiliation{Karlsruher Institut f\"ur Technologie, Institut f\"ur Festk\"orperphysik, Hermann-v.-Helmholtz-Platz 1, 76344 Eggenstein-Leopoldshafen, Germany}

\author{M. Minola}
\email{m.minola@fkf.mpg.de}
\affiliation{Max-Planck-Institut f\"ur Festk\"orperforschung, Heisenbergstrasse~1, 70569 Stuttgart, Germany}

\author{B. Keimer}
\email{b.keimer@fkf.mpg.de}
\affiliation{Max-Planck-Institut f\"ur Festk\"orperforschung, Heisenbergstrasse~1, 70569 Stuttgart, Germany}

\date{\today}

\begin{abstract}
We have used high-resolution resonant inelastic x-ray scattering (RIXS) to study a thin film of NdNiO$_3$,  a compound whose unusual spin- and bond-ordered electronic ground state has been of long-standing interest.
Below the magnetic ordering temperature, we observe well-defined collective magnon excitations along different high-symmetry directions in momentum space.
The magnetic spectra depend strongly on the incident photon energy, which we attribute to RIXS coupling to different local electronic configurations of the expanded and compressed NiO$_6$ octahedra in the bond-ordered state.
Both the noncollinear magnetic ground state and the observed site-dependent magnon excitations are well described by a model that assumes strong competition between the antiferromagnetic superexchange and ferromagnetic double-exchange interactions.
Our study provides direct insight into the magnetic dynamics and exchange interactions of the rare-earth nickelates, and demonstrates that RIXS can serve as a site-selective probe of magnetism in these and other materials.
\end{abstract}

\maketitle

\section{Introduction}

The spin ordering and dynamics in transition metal oxides are of long-standing interest from both fundamental and practical points of view~\cite{Hwang2012,Scalapino2012}.
Recent years have witnessed a surging effort in realizing a new generation of spintronic devices that may outperform the conventional electronic ones using magnetic oxides~\cite{Bibes2007,Bea2008,Bibes2011}.
Many of these proposals are based on artificial  heterostructures~\cite{Yamada2004}, or materials including multiple magnetic elements such as the double perovskites~\cite{Kobayashi1998}.
A key issue for a fundamental understanding of the correlated-electron physics at play in these complex materials, and consequently for a rational design and realization of oxide-based devices, is how to selectively probe the electronic and magnetic properties of the different functional components.

RIXS has recently emerged as a versatile tool for probing the ordering and dynamics of spin, charge, orbital, and lattice degrees of freedom in solids~\cite{Ament2011,Braicovich2010,LeTacon2011,Dean2013,Lee2014,Schlappa2012,Kim2014,Fabbris2017}.
The chemical selectivity endowed by its resonant nature enables RIXS to individually address the properties that stem from different active elements, as well as their interactions in complex materials.
In addition, the large cross section at x-ray resonance edges combined with the high photon flux at modern synchrotron radiation sources allows RIXS experiments on films and heterostructures with thickness down to a single unit cell~\cite{Dean2012,Ghiringhelli2014}.
Because of limitations in energy resolution, soft x-ray RIXS experiments have thus far mostly focused on cuprate compounds due to their exceptionally large magnon bandwidth.
However, the latest advances in RIXS instrumentation now allow detection of collective magnetic excitations in a broader class of metal-oxides~\cite{Betto2017,*Chiuzbaian2005,*Ghiringhelli2009}.

In this article, we take advantage of the emerging high-resolution soft x-ray RIXS capabilities to study the magnons in the perovskite rare-earth ($R$) nickelates \RNO{} which exhibit an intriguing phase diagram controlled by the radius of $R$~\cite{Torrance1992,GarMun1994,Catalan2008}.
Except for LaNiO$_3$, bulk \RNO{} undergo a metal-insulator transition (MIT), accompanied by breathing-type lattice distortions forming a rock-salt pattern of expanded (Ni$_A$) and compressed (Ni$_B$) NiO$_6$ octahedra~\cite{Catalan2008} [see Fig.~\ref{fig:ov}(a)].
It was suggested~\cite{Zhou2000} that this transition is an order-disorder transition of the Ni-O bond-length fluctuations pre-existing in the metallic phase. Concerning the nature of the MIT, various factors such as Fermi-surface nesting~\cite{Lee2011,Lu2017}, Hund's coupling~\cite{Mazin2007}, and strong electron-lattice coupling~\cite{Medarde1998,Jaramillo2014,Littlewood2017} have been discussed.
The particular importance of small/negative charge-transfer gap physics~\cite{Zaanen1985} for the bond-disproportionation in bulk \RNO{} has been emphasized~\cite{Johnston2014,Green2016,Park2012,Park2014,Lau2013,Subedi2015}.

The magnetism of insulating \RNO{} is unusual.
The Ni$_{A(B)}$ sites with more ionic (covalent) Ni-O bonds host spins of different sizes with $S_A>S_B$~\cite{Fernandez2001}.
At low temperatures, they condense into a four-sublattice noncollinear antiferromagnetic (AF) state with an ordering vector $\vb{Q}=(1/4,\,1/4,\,1/4)$ (in pseudocubic notation).
The nearest-neighbor $S_A$ and $S_B$ spins are approximately orthogonal to each other~\cite{Fernandez2001,Scagnoli2006,Scagnoli2008,Frano2013}, and the magnetic order can thus be viewed as a 90$^\circ$ spin spiral.
Such an ordering has no analogue in insulating oxides, and its underlying mechanism remains a puzzle.
One of the main obstacles to understanding the magnetism of \RNO{} lies in the difficulties related to the characterization of different local spin states of the two inequivalent Ni sites and their associated exchange interactions.
Although the two Ni sublattices are expected to respond differently in magnetic susceptibility~\cite{Park2012}, x-ray absorption~\cite{Green2016}, and optical spectroscopy~\cite{Ruppen2015} measurements, such predictions are difficult to test in practice, as these methods measure only the averaged properties of both sublattices.

In this work we have employed the unique site-selective capabilities of RIXS to probe the low-energy spin excitations associated with the two Ni sublattices in \NNO{}.
Well-defined magnon dispersions along different high-symmetry directions in the first Brillouin zone were observed and reproduced by model calculations.
After decades of research on the electronic phase behavior of \RNO, these experiments finally determine the magnetic exchange interactions and hence yield unprecedented insight into the mechanism driving the formation of the unusual noncollinear magnetic structure in this system.

\section{Experimental Details and Results}

The RIXS experiment was performed at the ID32 beamline of the European Synchrotron Radiation Facility using the new ERIXS spectrometer~\cite{erixs}.
We varied the scattering angle in the range from 55 to 135 degrees, which corresponds to momentum transfer of 0.4 to 0.8 \AA$^{-1}$ at the Ni $L_3$ edge around 853~eV.
The combined instrumental energy resolution was set to $\sim$50~meV [full width at half maximum (FWHM)] to allow for reasonable acquisition time at sufficiently high resolving power.
The incident photon polarization was kept parallel to the scattering plane.
The 40~nm thick \NNO{} film was grown on a [001]-oriented SrTiO$_3$ substrate by pulsed laser deposition.
It shows \emph{bulk-like} ordering behavior with transition temperature \tmi{}\,$\approx$~200~K~\cite{Frano2013,Lu2016,SupMat}. The average Ni-O bond disproportionation in the insulating phase was determined to be comparable to the value found in the bulk~\cite{Lu2016}.

\begin{figure}[tb]
  \includegraphics{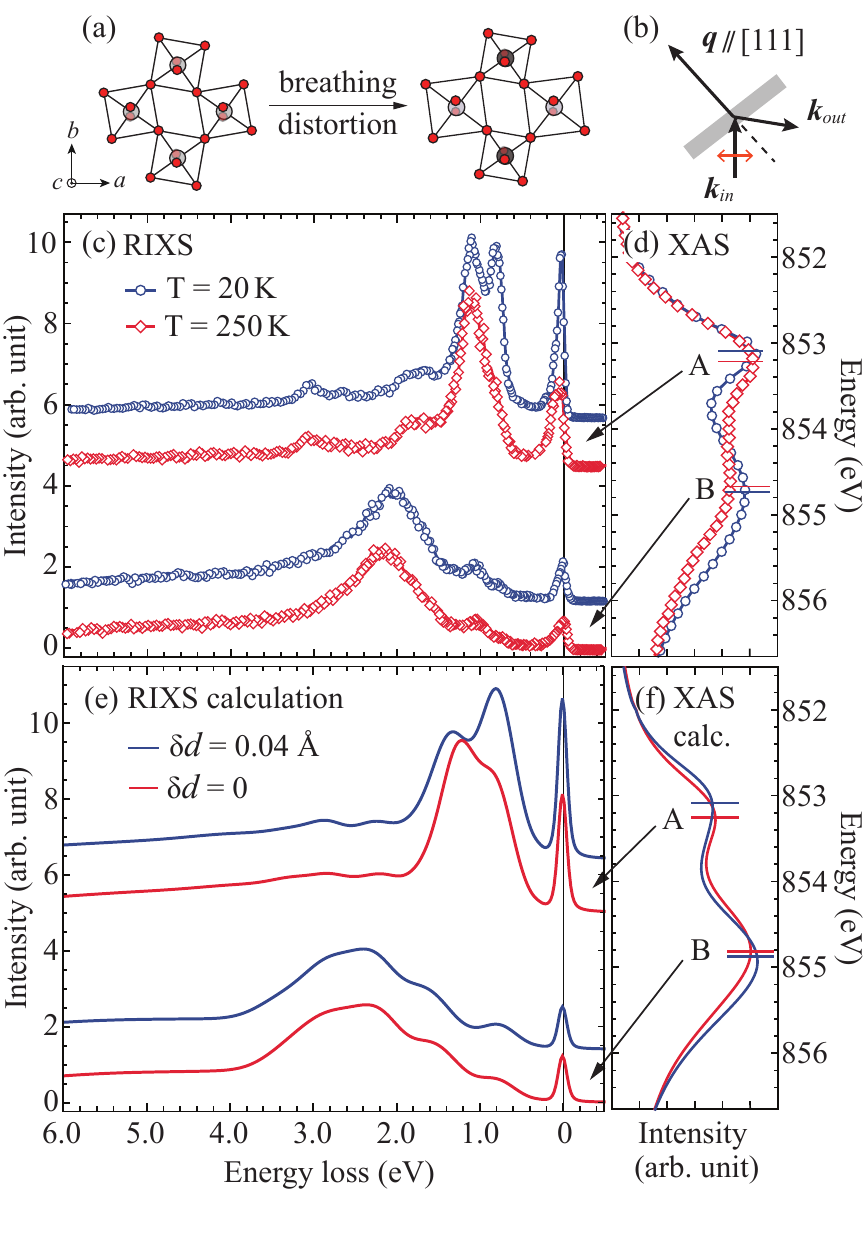}
  \caption{\label{fig:ov} (a) Sketch of the breathing lattice distortion across the MIT in the \RNO. (b) The scattering geometry used in the experiment. (c) RIXS spectra and (d) XAS for the \NNO{} film at temperatures below and above \tmi. The RIXS spectra were collected at $q_{111}=0.194$ with incident photon energy tuned to the resonant peaks $A$ and $B$ marked by bars in panel (d). (e) RIXS spectra and (f) XAS calculated using a double-cluster model with $\delta d=$ 0.04 \AA{} (blue) and 0 \AA{} (red) with the experimental geometry and light polarization in (c) and (d).}
\end{figure}

Figures~\ref{fig:ov}(c) and (d) show the RIXS and electron-yield x-ray absorption spectra (XAS) measured at the Ni $L_3$ edge ($2p\rightarrow3d$) on the \NNO{} film at temperatures both below and above \tmi.
The scattering geometry is shown in Fig.~\ref{fig:ov}(b).
The two sets of RIXS spectra shown in Fig.~\ref{fig:ov}(c) were collected at momentum transfer $q_{111}=0.194$ [shorthand for $\vb{q} = 0.194(1,1,1)$] with the incident photon energy $E_i$ tuned to that of the two resonant peaks $A$ and $B$ around $E_A\approx$ 853 eV and $E_B\approx$ 855~eV in the XAS [Fig.~\ref{fig:ov}(d)], respectively.
Similarly to a previous study~\cite{Bisogni2016}, the high-energy parts (with energy loss $\omega=E_i-E_f > 0.3$~eV) of the two datasets are very different.
Given that the RIXS intermediate states are the same as the XAS final states, it is helpful to examine the nature of the two peaks A and B in the XAS.
As \RNO{} is believed to be ``self-doped''~\cite{Mizokawa1991,Mizokawa2000} with a ground state configuration of predominantly $3d^8\underline{L}$ character (where $\underline{L}$ denotes a ligand hole), the sharp peak A can be assigned to the final state $2\underline{p}\,3d^9\underline{L}$ and the broad peak B to $2\underline{p}\,3d^8$, which correspond to excitations of 2$p$ core electron into the $3d$ and ligand states, respectively~\cite{Green2016}.
Based on this assignment, the decay process of RIXS with $E_i=E_A$ involves mostly transitions of the type $2\underline{p}\,3d^9\underline{L}\rightarrow (3d^8)^*\underline{L}$, which give rise to large $d$-$d$-excitation spectral weight centered around energy loss of 1~eV [upper part of Fig.~\ref{fig:ov}(c)].
For spectra measured with $E_i=E_B$, on the other hand, the core hole is annihilated largely by the ligand electrons, resulting in the broad fluorescence peak around 2-3 eV [bottom part of Fig.~\ref{fig:ov}(c)].
In both cases the RIXS final states also include $3d^9\underline{L}^2$ and associated particle-hole excitations, which generate a broad charge-transfer background.
Fig.~\ref{fig:ov}(e) and (f) show the RIXS spectra and XAS calculated with multiplet ligand-field theory~\cite{Haverkort2012,Lu2014,Haverkort2014} using a double-cluster model~\cite{Green2016}. The spectra calculated with bond disproportionation $\delta d$ set to 0 \AA (0.04 \AA~\cite{Lu2016}) correspond to those measured at 250~K (20~K), respectively. At both temperatures, the main features discussed above are well captured by the calculation. The details are discussed in Appendix~\ref{sec:cluster}.

From the above observations, it follows that in the bond-disproportionated state with charge redistribution of the form $(3d^8\underline{L})(3d^8\underline{L})\rightarrow (3d^8)_A(3d^8\underline{L}^2)_B$~\cite{Johnston2014,Green2016},
the distinct local configurations of the Ni$_A$ and Ni$_B$ sites give rise to remarkably different energy profiles of their absorption spectra, as shown in Appendix~\ref{sec:cluster}.
For the Ni$_A$ (Ni$_B$) sites of majority $3d^8$ ($3d^8\underline{L}^2$) character, the XAS arises dominantly from excitations into the unoccupied 3$d$ (ligand) states with spectral weight centered around energy A (B) [Fig.~\ref{fig:ov}(d)].
A similar site-dependent behavior is also seen for the x-ray magnetic circular dichroism (XMCD)~\cite{Green2016}.
By recalling that the coupling of low-energy magnon excitations to RIXS is controlled by XMCD~\cite{Haverkort2010}, one realizes that the spin excitations associated with the two Ni sites can be individually probed by selectively tuning the incident photon energy to their respective resonances.

\begin{figure}[tb]
  \includegraphics{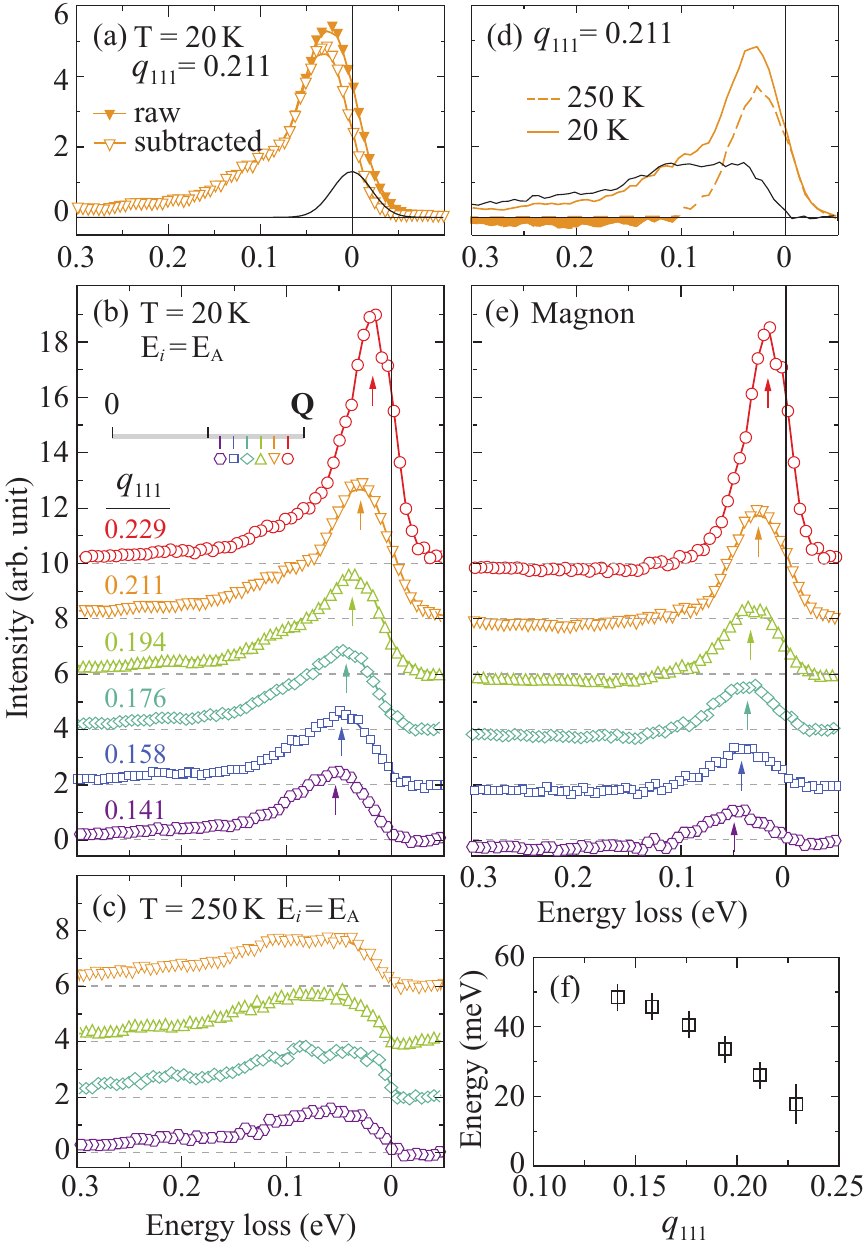}
  \caption{\label{fig:le111} (a) An example of the low-energy RIXS spectra of the \NNO{} film measured with $E_i=E_A$ at $q_{111}=0.211$ at 20~K. The elastic contribution is fitted by a Gaussian profile at $\omega=0$ with FWHM$~\sim$50~meV. (b) and (c) Spectra at different $q_{111}$ values after subtracting the elastic part at (b) 20 K and (c) 250 K. The inset in panel (b) sketches the momentum transfer $\vb{q}$ of the spectra with respect to the ordering vector $Q_{111}=0.25$. (d) The magnon scattering intensity can be obtained by subtracting the non-dispersing contribution (c) from the inelastic spectra in panel (b), as illustrated at $q_{111}=0.211$. The resulting spectra at all measured $q_{111}$ values are shown in panel (e). (f) The magnon dispersion obtained from panel (e).}
\end{figure}

In the following, we focus on the RIXS spectra below 0.3~eV, which show clear asymmetric lineshapes arising from low-lying excitations, as exemplified in Fig.~\ref{fig:le111}(a).
To extract the low-energy excitations, the elastic contribution can be removed by subtracting a Gaussian peak at $\omega=0$ with FWHM set to the experimental energy resolution of $\sim$ 50~meV.
The subtraction is performed for all the spectra presented hereafter.
The raw spectra and the subtraction procedure can be found in the Supplemental Material~\cite{SupMat}.
Figure~\ref{fig:le111}(b) shows an overview of the RIXS spectra measured at 20~K along the $[111]$ momentum-transfer direction.
The incident photon energy is tuned to $E_i=E_A$, which is expected to couple dominantly to the spins of Ni$_A$ sites.
The $\vb{q}$ values range from 0.40 to 0.65~\AA$^{-1}$ in steps of 0.05~\AA$^{-1}$, which correspond to $q_{111} =$ 0.141 to 0.229.
A clear inelastic contribution peaked at $\omega\lesssim 50$~meV can be observed at all measured $\vb{q}$ values.

We employ several complementary diagnostic indicators to discriminate between magnon and phonon contributions to the inelastic intensity. First, as magnons become heavily overdamped (and hence nearly invisible to scattering probes) in the paramagnetic state, we compare RIXS spectra well above (Fig.~\ref{fig:le111}(c)) and well below (Fig.~\ref{fig:le111}(b)) the antiferromagnetic ordering temperature. In the paramagnetic state, the spectra exhibit no noticeable $\vb{q}$ dependence, as expected for optical phonons~\cite{Devereaux2016,Hepting2014}. In addition, multimagnons~\cite{Bisogni2012} and charge excitations~\cite{Devereaux2016} may also contribute to these featureless spectra~\footnote{We note, however, that the charge excitations in RIXS are expected to be highly momentum dependent, and thus their contribution to the dispersionless spectra should be small.}. In contrast, the spectra in the AF state exhibit a prominent, strongly dispersive feature [indicated by arrows in Fig.~\ref{fig:le111}(b)]. The dispersion emanates from the AF ordering wavevector $\vb{Q}$ and exhibits maxima at the borders of the AF Brillouin zone, as expected for collective excitations of the AF state. In addition, the intensity is maximal at $\vb{Q}$ and minimal at the AF Brillouin zone border, again supporting the assignment to magnons. The collective nature of the magnetic excitation is further verified by the independence of its energy on variation of $E_i$ around $E_A$~\cite{SupMat}. We emphasize that the unique energy selectivity of RIXS allowed us to maximize the cross section of local excitations by utilizing its relatively large energy separation from the continuum excitations in the RIXS intermediate state.

We proceed to extract the magnetic RIXS spectra by subtracting the non-dispersing background measured at 250~K from the spectra at 20~K, as illustrated in Fig.~\ref{fig:le111}(d) for $q_{111}=0.211$. We notice the presence of a small negative intensity at energies above 0.1~eV in the subtracted spectra: this is most likely the result of additional particle-hole continuum emerging in the high-temperature metallic phase.
Figure~\ref{fig:le111}(e) shows the magnetic RIXS spectra at each $q_{111}$ after the subtraction.
The magnon excitation energies are obtained from the peak positions of the resultant spectra and plotted in Fig.~\ref{fig:le111}(f), where they exhibit a maximum of $\sim$50~meV at $q_{111} = 0.141 \approx \vb{Q}/2$. This energy scale of the magnetic dynamics is also confirmed by spectra measured along the [101] momentum-transfer direction~\cite{SupMat}.

\begin{figure}[tb]
  \includegraphics{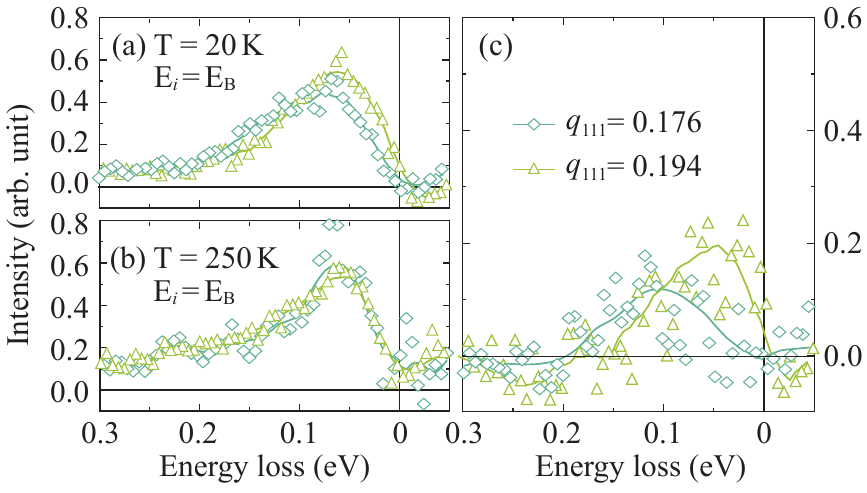}
  \caption{\label{fig:le111sat} Low-energy part of the RIXS spectra measured with $E_i=E_B$ at (a) 20 K and (b) 250 K after the subtraction of the elastic contribution. Panel (c) shows the magnon scattering intensity obtained by using the same procedure as in Fig.~\ref{fig:le111}(c). The lines are a guide to the eye.}
\end{figure}

Now we turn to the low-energy spectra measured with $E_i=E_B$, i.e., the energy that maximizes the RIXS coupling to the magnon excitations associated with the Ni$_B$ sites.
Figures~\ref{fig:le111sat}(a) and \ref{fig:le111sat}(b) show the spectra measured along [111] at 20~K and 250~K, respectively~\footnote{We note that the two data points of intensity 0.8 at 250 K are considered as noise. They cannot be an intrinsic feature of the spectrum, because the width of the corresponding peak would be 15 meV, well below our instrumental resolution of 50 meV}.
Contrary to the data taken with $E_i=E_A$, where the RIXS signal gains significant weight from collective magnon excitations, the spectra here are dominated by the non-dispersing contribution.
Following the same procedure as in Fig.~\ref{fig:le111}(d), we subtract the 250~K spectra from the 20~K ones and plot in Fig.~\ref{fig:le111sat}(c) the resulting spectra that one can assign to magnon excitations.
While the remaining spectral weight may indicate a small contribution from dispersive magnons, the signal is too weak to be clearly distinguished from the statistical noise.
The well-defined dispersing magnon excitations observed at $E_i=E_A$ and the nearly null observation at $E_i=E_B$ are in line with the prediction of the site-resolved spin susceptibilities~\cite{Park2012}.

\section{Magnetic RIXS calculation}

\begin{figure}[tb]
  \includegraphics{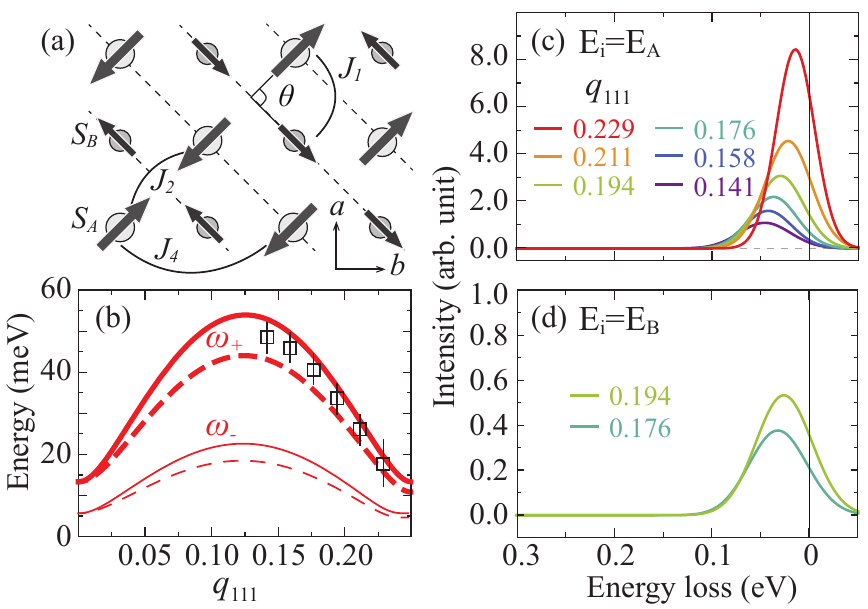}
  \caption{\label{fig:sw} (a) Schematic of the magnetic structure of \RNO{}. It can be viewed as stacking of alternating ferromagnetic $S_A$ and $S_B$ planes along [111], as indicated by the dashed lines. (b) Calculated spin-wave dispersion along [111] for two kinds of domains (solid lines for $hkl$ and $\bar{h}\bar{k}l$, and dashed for $h\bar{k}l$ and $\bar{h}kl$), each includes two branches (thick and thin) due to the disproportionated spins. Open symbols are experimental data from Fig.~\ref{fig:le111}(f). (c) and (d) Calculated RIXS spectra of magnon excitations for the experimental geometry at $E_i=E_A$ and $E_B$.}
\end{figure}

The combined knowledge of the static AF ordering and the collective excitations of the Ni moments provides a solid basis --- and simultaneously imposes stringent constraints --- for low-energy spin models aimed to describe the hitherto unexplained noncollinear magnetic structure
of \RNO{}~\cite{Scagnoli2006,Scagnoli2008,Frano2013}.
Figure~\ref{fig:sw}(a) shows a top view of the magnetic structure in \RNO{}, where ferromagnetic (111) planes form a noncollinear ($\uparrow \leftarrow \downarrow \rightarrow$) AF order.
The ordering pattern of spins $S_A$ ($S_B$) residing on the Ni$_A$ (Ni$_B$) sublattice readily follows from AF superexchange (SE) interactions $J_2$ and $J_4$, provided that $J_4>J_2/2$.
The latter condition can be justified by the large $pd\sigma$ charge-transfer fluctuations along the straight Ni-O-Ni-O-Ni bonds associated with $J_4$~\cite{KimJH2014}.
However, the major question is which mechanism stabilizes the $90^{\circ}$ mutual orientation of the $S_A$ and $S_B$ sublattices, given that a conventional (Heisenberg) coupling $({\bf S}_A{\bf S}_B)$ is frustrated and would result (via the ``order-by-disorder'' mechanism) in a collinear ``up-up-down-down'' arrangement instead.

We suggest that, near the MIT, low-energy charge fluctuations between Ni$_A$ and Ni$_B$ sites with unequal spins $S_A>S_B$ lead to a double-exchange (DE) process that depends on the angle $\theta$ between $S_A$ and $S_B$ as $-\cos\frac{\theta}{2}$~\cite{Zener1951,Anderson1955,deGennes1960}. Competition between ferromagnetic DE and AF superexchange is known to result in noncollinear ordering~\cite{deGennes1960} as recently observed in ferrates~\cite{KimJH2014}.
In the present context, the DE energy gain from the ``left'' and ``right'' $AB$-bonds in each direction [Fig.~\ref{fig:sw}(a)] is proportional to $-(\cos \frac{\theta}{2} + \cos \frac{\pi-\theta}{2})$, which is optimized at $\theta=\pi/2$ and thus supports an unusual $90^{\circ}$ alignment of $S_A$ and $S_B$.

As far as small fluctuations around the classical spin pattern are concerned, one can incorporate the DE energy into a spin-only model~\cite{Khaliullin2000,KimJH2014} with an effective parameter
$J_1=J_{SE}-J_{DE}$ comprising both SE and DE contributions, where $J_{DE}$ is associated with the kinetic energy due to low-energy charge fluctuations between Ni$_A$ and Ni$_B$ sites.
We have calculated the magnon dispersions using the $J_1$-$J_2$-$J_4$ model, including also a single-ion anisotropy term $K$ assuming easy axes parallel to the cubic body diagonals.
The spin disproportionation, quantified as $S_{A(B)}=\frac{1}{2}(1\pm\delta)$, results in two magnon branches, $\omega_+$ and $\omega_-$, with the former one carrying most of the scattering intensity.
The disproportionation parameter $\delta = 0.4$ was chosen such that it gives $S_A/S_B \approx 2$, consistent with experiment~\cite{Fernandez2001} and with the prediction of the double-cluster model for the experimental value of the bond disproportionation~\cite{Green2016} (Fig.~\ref{fig:ov} and associated discussion). Details of the calculations are presented in Appendix~\ref{sec:model}.

The spin Hamiltonian illustrated by Fig. ~\ref{fig:sw}(a) contains four adjustable parameters: $J_1$, $J_2$, $J_4$, and $K$. We note that the major magnon branch $\omega_+$ is not sensitively dependent on the exact value of $J_1$. An accurate determination of $J_1$ requires resolution of the less intense, lower-energy branch $\omega_{-}$ which is beyond the capabilities of the current setup. However, the experimentally observed commensurate ordering wavevector implies the constraint $|J_1|< J_{1\mathrm{c}}=2 \sqrt{2K(J_2+J_4)/3}$ (Appendix~\ref{sec:model}). In the fitting procedure, $J_1$ was therefore varied systematically between its upper and lower bounds to establish systematic errors for $J_2$, $J_4$, and $K$. Note that the maximum and minimum of $\omega_+$ along the [111] direction and its maximum along the [101] direction~\cite{SupMat} constitute three well-defined, independent pieces of experimental information that allow us to obtain these three parameters with good accuracy. The additional data points at different wavevectors further increase the confidence in these parameters.

Together with the statistical errors arising from the experimental uncertainty in determining the magnon energies, we obtain the following values and errors for the fitting parameters:  $J_2=  4.1 \pm 1.7$,   $J_4 = 8.3 \pm 2.2$, and $K = 1.3 \pm 1.3$ meV, which implies $J_{1\mathrm{c}} = 6.3$ meV. Fig. ~\ref{fig:sw}(b) shows the magnon dispersions calculated with the best-fit parameters and $J_1 =  J_{1\mathrm{c}}/2$. The fitted value of the anisotropy constant $K$ results in a magnon gap of 13 meV for the main branch of the dispersion, which is comparable to that observed in La$_2$NiO$_4$~\cite{Nakajima1993}. However, the magneto-crystalline anisotropy is immaterial for our model and interpretation, which addresses the hierarchy of exchange interactions between the Ni ions. We emphasize once more that the dominance of the long-range interactions, $J_2$ and $J_4$, is a hallmark of the competition between superexchange and double-exchange interactions near the MIT, which greatly reduces the nearest-neighbor interaction $J_1$.

To calculate the magnetic spectra probed by RIXS, we use the RIXS operator~\cite{Haverkort2010} $R_{A(B)}=\sigma^{(1)}_{A(B)}(\bm{\epsilon}_i \times \bm{\epsilon}_o^*)\cdot \vb{S}_{A(B)}$ with
$\bm{\epsilon}_{i(o)}$ the polarization of the incident (outgoing) photons.
$\sigma^{(1)}_{A(B)}$ are the calculated site-resolved XMCD spectra for the Ni$_A$ (Ni$_B$) sites.
The calculated magnetic RIXS spectra for $E_i=E_A$ and $E_B$ are plotted in Figs.~\ref{fig:sw}(c) and \ref{fig:sw}(d), respectively.
All the spectra are averaged over contributions from the four $hk$-domains and then
scaled with a universal factor to match the experimental intensity.
For incident photon energy $E_i=E_A$, excellent agreement is achieved between calculation and experiment for the energy and spectral weight of the magnon excitation at all measured $\vb{q}$ values (including $\vb{q}\parallel$[101]~\cite{SupMat}), providing strong support for the proposed spin model. For $E_B$, while both calculation and experiment show strongly suppressed magnon spectral weight, the calculation shows a small deviation from the measurement which could be attributed to the experimental sensitivity at its limit and/or to quantum effects beyond the linear-spin-wave approximation.
While a state with nonzero moments on site A only cannot be rigorously ruled out, the RIXS results and their theoretical description are quantitatively consistent with a non-collinear magnetic state with different but nonzero moments on sites A and B.

\section{Summary}

In summary, we have studied the electronic and magnetic excitations in \NNO{} with RIXS at the Ni $L_3$ edge.
The observed complex energy dependence of the RIXS spectra is attributed to the different local configurations and spin states in the bond-disproportionated state and their consequent distinct coupling to the RIXS process.
The observed magnetic order and excitations are well captured by an effective spin model based on strong competition between superexchange and double-exchange interactions in \RNO. Our study demonstrates the capability of RIXS to determine site-selective contributions to the collective magnetic dynamics, and provides an example for its application in oxide thin films and heterostructures that contain several functional
materials and/or inequivalent structural sites. With the development of future x-ray sources and other high-resolution RIXS setups, such as the one that was recently installed at beamline I21 of the Diamond Light Source~\cite{I21}, we expect that these advances will come to full fruition.

\begin{acknowledgments}
We are grateful to G. Ghiringhelli and L. Braicovich for facilitating the RIXS measurements using the ERIXS spectrometer.
We thank E.~Lefran\c{c}ois for assistance at the ID32 beamline.
\end{acknowledgments}

Y. L. and D. B. contributed equally to this work.

\appendix

\section{XAS and RIXS spectra calculated using a double-cluster model}\label{sec:cluster}


The XAS and RIXS spectra shown in Fig.~\ref{fig:ov}(e) and (f) were calculated using a double-cluster model as proposed in Ref.~\cite{Green2016}, which includes two NiO$_6$ clusters with inter-cluster coupling between the Ni-$d$ and ligand states. The details of the model construction can be found in Ref.~\cite{Green2016}.

The parameters for the single-particle part of the Hamiltonian in our calculation are derived from downfolded values using the Stuttgart-$N$MTO program~\cite{Andersen2000,nmto}. Following the same convention of Ref.~\cite{Green2016}, we define the parameters in the non-disproportionated state as $V_{eg}=2.8$, $V_{t2g}=1.5$, 10$Dq$ = 0.9, and $T_{pp}=0.5$, all in units of eV. The monopole Coulomb interaction within the $d$ shell is set to $U_{dd}=6.0$ eV, and between the $d$ and core $p$ shells to $U_{pd}=7.0$ eV. The multipole part is set to 80\% of the Hartree-Fock values of the Slater integrals~\cite{Freeland2016}. The charge-transfer energy is $\Delta=-1.0$ eV. For the bond-ordered state, the disproportionation value $\delta d= d^A_{\text{Ni-O}}-d^B_{\text{Ni-O}}$ is defined as the deviation of the average bond length $d^{A(B)}_{\text{Ni-O}}$ of the expanded A (compressed B) NiO$_6$ octahedra from their mean value. For the \NNO{} film used in this study $\delta d=0.04$~\AA~\cite{Lu2016}. The inter-cluster mixing is set to 0.33~\cite{Green2016} to reproduce the experimental energy splitting between A and B.

We note that the relative intensity of the A resonance with respect to B in the experiment is slightly higher than that in the calculation as well as that observed in previous studies~\cite{Medarde1992}, which may be an indication of an additional Ni$^{2+}$ contribution most likely from oxygen off-stoichiometry at the sample surface.

For the RIXS spectra measured at A, the $dd$ excitation around 1~eV and its change of line-shape across the bond ordering transition is well reproduced. For the RIXS measured at B, the fluorescence intensity around 2-3 eV dominates the spectral weight, similar to the experimental observation. For both incident energies, the RIXS spectra show broad charge-transfer spectral weight centered around 4~eV, again in good agreement with the experimental data.


\begin{figure}[h]
  \includegraphics{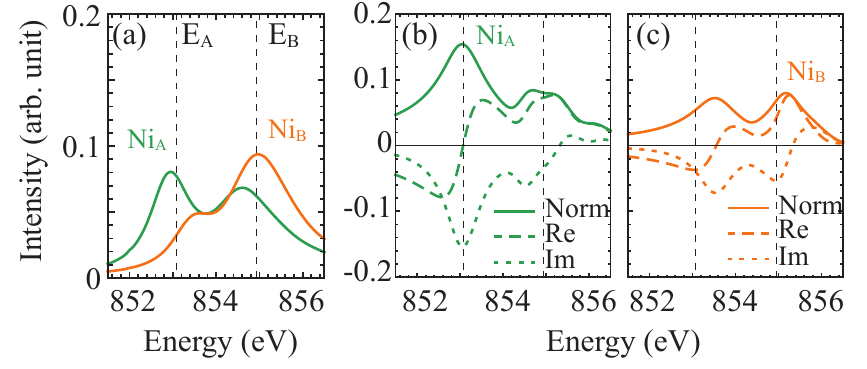}
  \caption{\label{fig:xmcd} Site-resolved (a) XAS and (b, c) XMCD spectra $\sigma^{(1)}$ calculated using the double-cluster model. The spectra for the expanded and compressed Ni sites are plotted in green and orange, respectively. The two vertical dashed lines mark the energies $E_A$ (left) and $E_B$ (right) within each panel.}
\end{figure}

Figure~\ref{fig:xmcd} shows the site-resolved XAS and XMCD spectra calculated using the above double-cluster model. The energies $E_A$ and $E_B$ are marked with vertical dashed lines in each panel. For the XAS, the expanded and compressed Ni sites contribute dominantly to peaks A and B, respectively. The XMCD spectra show a similar site-dependence as the XAS. The energy-dependent coupling to spin excitations in RIXS can be estimated by the norm of the XMCD spectra, which shows a maximum close to $E_A$ and $E_B$ for the expanded and compressed sites, respectively.

\section{Magnetic Model and Spin-Wave Calculation}\label{sec:model}

Considering small fluctuations of spins near the classical configuration in Fig.~\ref{fig:sw}(a), we derived the magnon dispersions within the $J_1$-$J_2$-$J_4$ model, supplemented by a single-ion anisotropy term $-KS_\alpha^2$, where $\alpha$ refers to the easy axis directions assumed to be oriented along the cubic body diagonals (as suggested by experiments~\cite{Fernandez2001,Scagnoli2006,Scagnoli2008,Frano2013}). The result reads as
\begin{equation}
  \omega_\pm(\bm{q}) = 3 \sqrt{X_{\bm{q}}Y_{\bm{q}}[(1+\delta^2) \pm R_{\bm{q}}]}
\end{equation}
where the subscript $+$ ($-$) denotes the upper (lower) branch of the dispersion.
The shorthands here are
\begin{equation}
  \begin{split}
    X_{\bm{q}} & = 2J_2 \eta_{\bm{q}} + J_4(1-\gamma_{2\bm{q}}) + K/3 \\
    Y_{\bm{q}} & = 2J_2 \xi_{\bm{q}} + J_4(1+\gamma_{2\bm{q}}) + K/3, \\
    R_{\bm{q}} & = \sqrt{(2\delta)^2 + \lambda^2_{\bm{q}}}, \\
  \end{split}
\end{equation}
with
\begin{equation}
  \begin{split}
    \eta_{\bm{q}} & = \frac{1}{3} (\sin q_x \sin q_y + \sin q_y \sin q_z + \sin q_z \sin q_x), \\
    \xi_{\bm{q}} & = \frac{1}{3} (\cos q_x \cos q_y + \cos q_y \cos q_z + \cos q_z \cos q_x), \\
    \gamma_{\bm{q}} & = \frac{1}{3}(\cos q_x + \cos q_y + \cos q_z), \\
    \lambda_{\bm{q}} & = (1-\delta^2) \frac{J_1 \gamma_{\bm{q}}}{Y_{\bm{q}}}.
  \end{split}
\end{equation}
At the $J_1=0$ limit, $\omega_+$ and $\omega_-$ correspond to the independent magnon dispersions within the $S_A$ and $S_B$ sublattices, respectively.
The stability condition $\omega_{\pm}>0$ requires $|J_1|< J_{1c}=2 \sqrt{2K(J_2+J_4)/3}$. For $|J_1| \geq J_{1c}$, the magnetic structure becomes incommensurate at $\bm{q} = \vb{Q}(1+\frac{1}{2\pi}\frac{J_1}{J_2+J_4})$.

To calculate the magnetic RIXS intensity, the RIXS operator is approximated as $R_{A(B)}=\sigma^{(1)}_{A(B)} (\bm{\epsilon}_i \times \bm{\epsilon}_o^*)\cdot \vb{S}_{A(B)}$ with $\bm{\epsilon}_{i(o)}$ the polarization of the incident (emitted) photons. The RIXS intensity for a given $\bm{q}$ can be then calculated as
\begin{equation}
  \begin{split}
    I_\pm(\bm{q}) = & \frac{\omega_\pm(\bm{q}+\vb{Q})}{3X_{\bm{q}+\vb{Q}}} \times \\
    & [ (|\alpha|^2p_x^2+|\beta|^2p_z^2) \pm \frac{2\delta}{R_{\bm{q}+\vb{Q}}} (|\alpha|^2p_x^2-|\beta|^2p_z^2) ]\\
  + & \frac{3X_{\bm{q}}}{\omega_\pm(\bm{q})} p_y^2\{(|\tilde \alpha|^2+|\tilde \beta|^2) \pm \\
    & \frac{1}{R_{\bm{q}}}[2(|\tilde \alpha|^2-|\tilde \beta|^2)\delta +(\tilde \alpha^*\tilde \beta+\tilde \alpha\tilde \beta^*)\lambda_{\bm{q}}]\}
  \end{split}
\end{equation}
for the two branches. Here $\tilde \alpha = \alpha (1+\delta)$ and $\tilde \beta = \beta (1-\delta)$. $\bm{p}=(p_x,p_y,p_z)$ is the orientation vector $\bm{\epsilon}_i \times \bm{\epsilon}_o^*$.  $\alpha$ and $\beta$ are the calculated (complex) values of $\sigma^{(1)}$ for site $A$ and $B$ for a given incident energy, as discussed in Section S2. For each $\bm{q}$ value, the RIXS intensity is calculated by averaging over the four $hk$-domains as
\begin{equation}
  \begin{split}
    \bar I_\pm(q_x,q_y,q_z) =
    \frac{1}{4}[& I_\pm(q_x,q_y,q_z)\\
    +& I_\pm(-q_x,q_y,q_z)\\
    +& I_\pm(q_x,-q_y,q_z)\\
    +& I_\pm(-q_x,-q_y,q_z)].
  \end{split}
\end{equation}
Finally, the intensity is averaged over emission polarizations $\bm{\epsilon}_o$ both parallel and perpendicular to the scattering plane.\\

%

\end{document}